\documentclass[11pt,a4paper]{article}
\usepackage[utf8]{inputenc}
\usepackage[T1]{fontenc}
\usepackage[english]{babel}
\usepackage{amsmath,amssymb,amsthm,amsfonts}
\usepackage{graphicx}
\usepackage{geometry}
\usepackage{hyperref}
\usepackage{xcolor}
\usepackage{booktabs}
\usepackage{enumitem}
\usepackage{physics}

\theoremstyle{definition}

\title{\textbf{A Gauge Theory of Turbulence:}\\
\large Higgs Mechanism for the Gribov Parameter,\\
Intermittency and the Anomalous Dimension of the Scalar Field}
\author{V. E. R. Lemes\footnote{email: verlemes@gmail.com\\\small \em Instituto de F\'\i sica, Universidade do Estado do Rio de
	Janeiro,\\
	\small \em Rua S\~{a}o Francisco Xavier 524, Maracan\~{a}, Rio de Janeiro - RJ,
	20550-013, Brazil}}

\date{August 13, 2026\\[2pt] \small (revised version v2: September 25, 2026)}

\begin{document}
\maketitle

\begin{abstract}
We formalize the gauge structure of the Navier--Stokes equation for incompressible fluids, interpreting $\nabla\cdot\mathbf{v}=0$ as a gauge fixing analogous to the Coulomb gauge in electromagnetism. We construct the Martin--Siggia--Rose action and its full BRST symmetry, introducing BRST doublets for the Gribov parameter $\gamma$ and the monodromy phase $\theta$. Through a Higgs mechanism, $\gamma$ acquires a vacuum expectation value $\gamma_{0}$, generating a mass scale for the vorticity and the ghosts; we compute the one-loop effective potential and analyze vacuum stability. Intermittency --- measured by the exponents $\zeta_{n}$ --- is described by fluctuations of the Higgs field $\sigma$ around the condensate, via a Gribov-inspired log-Poisson hierarchy that satisfies exactly $\zeta_{3}=1$ (Kolmogorov's four-fifths law). A one-parameter fit to the DNS exponents of Gotoh \emph{et al.} favors filamentary vorticity structures, $D_{f}=1.10\pm0.15$, while a two-parameter fit shows no significant improvement, the $(D_{f},\Delta)$ plane being degenerate. The one-loop anomalous dimension in $d=3$, $\Delta_{\sigma}\approx0.50$, supports the identification $D_{f}=2\Delta_{\sigma}$ within the fit uncertainty. The formalism unifies classical hydrodynamics with gauge theory and spontaneous symmetry breaking, opening the study of turbulence and intermittency to quantum field theory methods.
\end{abstract}

\tableofcontents
\newpage

\section{Introduction}

The Navier--Stokes equation for an incompressible, Newtonian, viscous fluid,
\begin{equation}\label{eq:NS}
\rho\left(\frac{\partial\mathbf{v}}{\partial t}+\mathbf{v}\cdot\nabla\mathbf{v}\right)=-\nabla p+\mu\nabla^{2}\mathbf{v}+\mathbf{f},
\end{equation}
subject to the constraint
\begin{equation}\label{eq:incompressibility}
\nabla\cdot\mathbf{v}=0,
\end{equation}
is the starting point of classical hydrodynamics. Condition \eqref{eq:incompressibility} is interpreted here as a \emph{gauge fixing}, analogous to the Coulomb gauge $\nabla\cdot\mathbf{A}=0$ in electromagnetism. Two infinite-dimensional group structures underlie this reading, and it is important to distinguish them from the outset. The first is the group of volume-preserving diffeomorphisms, $\mathrm{SDiff}(\Omega)$, whose Lie algebra is the space of divergence-free vector fields: it is the particle-relabeling symmetry of the Euler equation, which is the geodesic flow on $\mathrm{SDiff}(\Omega)$ with the $L^{2}$ metric \cite{Arnold1966,MarsdenWeinstein1983}, and it is responsible, through Noether's theorem, for Kelvin's circulation theorem. The second is the additive group of harmonic functions, which acts on the velocity field itself and plays the role of the residual gauge group once the condition \eqref{eq:incompressibility} is imposed. It is this second structure that generates the Gribov copies of the present theory.

The residual gauge transformation is
\begin{equation}\label{eq:gauge_residual}
\mathbf{v}\to\mathbf{v}+\nabla\chi,\qquad p\to p-\rho\left(\frac{\partial\chi}{\partial t}+\mathbf{v}\cdot\nabla\chi+\frac{1}{2}|\nabla\chi|^{2}\right),
\end{equation}
with the requirement that $\chi$ be harmonic:
\begin{equation}\label{eq:laplace_chi}
\nabla^{2}\chi=0.
\end{equation}
The non-trivial solutions of \eqref{eq:laplace_chi} are the \emph{Gribov copies}, which correspond to vortices in the fluid. A remark is in order here: on a periodic domain the only smooth harmonic functions are constants, and the residual orbit is trivial. Non-trivial copies require either boundaries or multivalued/singular harmonic functions. The physically relevant case is precisely the latter: a multivalued $\chi=(\Gamma/2\pi)\,\vartheta$, with $\vartheta$ the azimuthal angle, generates through \eqref{eq:gauge_residual} an elementary potential vortex of circulation $\Gamma$. The identification ``Gribov copies $=$ vortices'' is therefore understood in the sense of singular gauge transformations. The boundary and regularity conditions that make this precise are given in Section~2.1.

It is important to establish the exact status of the transformation \eqref{eq:gauge_residual}. Using the Lamb form of the advection term, $\mathbf{u}\cdot\nabla\mathbf{u}=\nabla(|\mathbf{u}|^{2}/2)-\mathbf{u}\times(\nabla\times\mathbf{u})$, and the fact that $\nabla\times(\mathbf{v}+\nabla\chi)=\boldsymbol{\omega}$, one shows that
\begin{equation}\label{eq:lamb_identity}
(\mathbf{v}+\nabla\chi)\cdot\nabla(\mathbf{v}+\nabla\chi)
=\mathbf{v}\cdot\nabla\mathbf{v}+\nabla\!\left(\mathbf{v}\cdot\nabla\chi+\tfrac{1}{2}|\nabla\chi|^{2}\right)-\nabla\chi\times\boldsymbol{\omega},
\end{equation}
identically. Since $\nabla^{2}(\mathbf{v}+\nabla\chi)=\nabla^{2}\mathbf{v}$ for harmonic $\chi$, the transformed field satisfies the Navier--Stokes equation with the transformed pressure of \eqref{eq:gauge_residual} and with the modified forcing $\mathbf{f}-\rho\,\nabla\chi\times\boldsymbol{\omega}$. This means that the transformation is an exact symmetry of the incompressibility constraint, of the linear (Stokes) part of the dynamics, and of the irrotational sector, but not of the full nonlinear equation: the obstruction is the solenoidal term $\nabla\chi\times\boldsymbol{\omega}$, which for a generic flow is not a pure gradient and therefore cannot be absorbed into the pressure. This is not surprising, as the vorticity itself is invariant under \eqref{eq:gauge_residual}, while its advection by $\mathbf{v}$ is not. What the transformation defines exactly is an \emph{orbit structure on the space of transverse fields}: two transverse fields with the same vorticity differ by a harmonic field, and the additive group of harmonic functions acts on the fibers of the curl map. The Faddeev--Popov operator of the condition \eqref{eq:incompressibility} along this orbit is the Laplacian, whose zero modes are precisely the copies (Section~2.1). In this way, the Gribov problem of the present theory is a property of the kinematical reconstruction of the velocity from the vorticity --- exactly as, in Yang--Mills theory, the copies are defined by the gauge condition and the orbit, not by the equations of motion. The physical non-triviality of the copies is guaranteed by Kelvin's theorem: the circulations that label them are constants of the inviscid motion. Finally, the obstruction $\nabla\chi\times\boldsymbol{\omega}$ has a direct physical meaning: it is the vortex-stretching content of the flow. The residual Abelian structure is exact in the linear regime and is broken by the same mechanism that sustains the turbulent cascade, a point that fits the phase picture of Section~9.

\subsection{Relation to previous work}

The application of field-theoretic methods to Navier--Stokes turbulence has a long history, with the MSR formalism \cite{MSR1973} as its systematic starting point. The symmetry content of the Navier--Stokes field theory has been analyzed in detail by Canet, Delamotte and Wschebor \cite{Canet2015}, who exhibited two gauge symmetries --- the time-gauged Galilean symmetry and a time-gauged shift symmetry of the response field --- and derived the associated Ward identities, including the K\'arm\'an--Howarth relation. Those gauge symmetries act on the dynamical fields of the MSR generating functional and are distinct from the structure explored here: we identify the \emph{incompressibility constraint itself} with a gauge condition, and the residual transformations \eqref{eq:gauge_residual} with the remaining gauge freedom.

Gauge-theoretic formulations of turbulence with spontaneous symmetry breaking have been considered independently in the recent work of Ref.~\cite{SO3Turbulence2026}, where an $SO(3)$ connection built from the specific angular momentum plays the role of the gauge field and a radial-velocity field plays the role of the Higgs, in a Georgi--Glashow-like construction. Our construction is complementary: the gauge structure is Abelian and stems from $\mathrm{SDiff}(\Omega)$, and the role of the Higgs field is played by the Gribov parameter itself.

On the field-theory side, our machinery is borrowed from the Gribov--Zwanziger framework of Yang--Mills theories \cite{Gribov1978,Zwanziger1989}: the horizon term, quadratic both in the gauge field and in the Gribov parameter, the BRST doublet formalism, and the dimension-two condensates that generate a dynamical mass scale \cite{Dudal2005,VandersickelZwanziger2012}. The novelty of the present work is the transplantation of this machinery to classical hydrodynamics, where the Gribov parameter acquires a direct physical interpretation as an inverse length scale of the turbulent state.

\subsection{Scope and assumptions}

It is important to state clearly what is derived and what is postulated in this work. \emph{(i)} The incompressibility condition is a physical constraint, with the pressure as its Lagrange multiplier; its reading as a gauge fixing is made precise at the level of the orbit structure generated by the residual transformations \eqref{eq:gauge_residual} --- not a claim that incompressible flows are redundant descriptions of a compressible theory, nor that \eqref{eq:gauge_residual} is a symmetry of the full nonlinear dynamics, which it is not [see eq.~\eqref{eq:lamb_identity}]. \emph{(ii)} The $\gamma$-sector of the action (Section~3) is \emph{postulated} as the minimal BRST-exact sector that closes on a Gribov-type gap equation and generates a mass scale for the vorticity; it is not derived from the Navier--Stokes equation. \emph{(iii)} The MSR functional integral lives in $3+1$ dimensional space-time, but the static, equal-time statistics relevant for structure functions are naturally computed in $d=3$ Euclidean dimensions; accordingly, the effective potential of Section~5 is evaluated in $d=4$ space-time, while the anomalous dimension of Section~8 is computed in $d=3$. \emph{(iv)} The identification $D_{f}=2\Delta_{\sigma}$ proposed in Section~8 is a conjecture whose numerical prefactors depend on the normalization conventions of the vertices; it should be read as a target for more precise calculations, not as an established result.

\section{The Martin--Siggia--Rose action and the BRST symmetry}

In the Martin--Siggia--Rose (MSR) formulation \cite{MSR1973}, one introduces the response field $\tilde{\mathbf{v}}$ and the functional action
\begin{equation}\label{eq:MSR}
S_{\mathrm{MSR}}=\int d^{4}x\,\Bigg[\tilde{\mathbf{v}}\cdot\bigg(\partial_{t}\mathbf{v}+\mathbf{v}\cdot\nabla\mathbf{v}-\nu\nabla^{2}\mathbf{v}+\frac{1}{\rho}\nabla p\bigg)-i\nu\,\tilde{\mathbf{v}}\cdot\nabla^{2}\tilde{\mathbf{v}}\Bigg].
\end{equation}
The variation with respect to $p$ enforces incompressibility; the variation with respect to $\tilde{\mathbf{v}}$ recovers the Navier--Stokes equation.

The BRST symmetry \cite{BRST1976} is defined by the nilpotent operator $s$ ($s^{2}=0$). The transformations for the MSR fields are
\begin{equation}\label{eq:BRST_MSR}
\begin{aligned}
s\,\mathbf{v}&=\tilde{\mathbf{v}},\qquad s\,\tilde{\mathbf{v}}=0,\\
s\,p&=0,\\
s\,\chi&=c,\qquad s\,c=0,\qquad s\,\bar{c}=\lambda,\qquad s\,\lambda=0,
\end{aligned}
\end{equation}
where $c$ and $\bar{c}$ are the Faddeev--Popov ghosts. In the space-time notation of the functional integral we also write the velocity as a 4-vector, $A_{\mu}(x)\equiv v_{\mu}(x)$, so that the first line of \eqref{eq:BRST_MSR} reads
\begin{equation}\label{eq:BRST_A}
sA_{\mu}=\tilde{v}_{\mu},\qquad s\,\tilde{v}_{\mu}=0.
\end{equation}
The notation $A_{\mu}$ will be used interchangeably with $\mathbf{v}$ whenever the gauge-theoretic analogy is in the foreground (Sections~3--5); no field content beyond the MSR multiplet is implied.

\subsection{The residual orbit: boundary conditions, regularity and the Faddeev--Popov operator}
\label{sec:orbit}

Before introducing the Gribov doublet, let us make precise the connection between the residual transformation \eqref{eq:gauge_residual}, the group $\mathrm{SDiff}(\Omega)$, and the Gribov copies. As stated in Section~1, $\mathrm{SDiff}(\Omega)$ acts on the vorticity by advection and is not the group generated by \eqref{eq:gauge_residual}: the gradient transformations form the additive group of harmonic functions, which is the kernel of the curl on transverse fields --- the redundancy in the reconstruction of $\mathbf{v}$ from $(\nabla\cdot\mathbf{v},\nabla\times\mathbf{v})$.

We consider a bounded domain $\Omega$ with the no-penetration condition $\mathbf{v}\cdot\mathbf{n}=0$ on $\partial\Omega$. Preservation of this condition under \eqref{eq:gauge_residual} requires
\begin{equation}\label{eq:bc_chi}
\nabla^{2}\chi=0\quad\text{in }\Omega,
\qquad
\frac{\partial\chi}{\partial n}=0\quad\text{on }\partial\Omega.
\end{equation}
On a simply connected domain, \eqref{eq:bc_chi} implies $\chi=\mathrm{const}$: the residual orbit is trivial and there are no copies --- the analogue of the absence of a Gribov problem for Abelian theories with trivial topology. Non-trivial copies exist in exactly two situations. The first is that of multiply connected domains: by the Hodge--de Rham theorem, the space of harmonic fields $\mathbf{h}$ with $\nabla\cdot\mathbf{h}=\nabla\times\mathbf{h}=0$ and $\mathbf{h}\cdot\mathbf{n}=0$ on $\partial\Omega$ has dimension $b_{1}(\Omega)$, the first Betti number, and the independent copies are labeled by the circulations $\Gamma_{a}=\oint_{C_{a}}\mathbf{v}\cdot d\mathbf{l}$ around the $b_{1}$ independent cycles --- equivalently, by multivalued harmonic functions $\chi$ with periods $\oint_{C_{a}}d\chi=\Gamma_{a}$. The second is that of singular harmonic functions: allowing $\chi\in C^{\infty}(\Omega\setminus\Sigma)$, harmonic on $\Omega\setminus\Sigma$, with $\Sigma$ a set of codimension $\geq2$ (vortex lines in $d=3$, points in $d=2$), the elementary copy is $\chi=(\Gamma/2\pi)\,\vartheta$, for which $\nabla\chi=(\Gamma/2\pi r)\,\mathbf{e}_{\vartheta}$ is smooth, divergence-free and curl-free away from the core, with $\nabla\times\nabla\chi=\Gamma\,\delta^{2}(\mathbf{x}_{\perp})\,\hat{\mathbf{z}}$. As regularity conditions we require $\nabla\chi\in L^{2}_{\mathrm{loc}}(\Omega\setminus\Sigma)$, so that the kinetic energy is finite up to the core region, and that the circulation $\Gamma_{C}=\oint_{C}\nabla\chi\cdot d\mathbf{l}$ be constant on homology classes of loops in $\Omega\setminus\Sigma$; the core singularity is regularized by viscosity, as in the Lamb--Oseen solution, so that the singular copy corresponds to a genuine Navier--Stokes vortex.

The Faddeev--Popov operator of the gauge condition \eqref{eq:incompressibility} along the residual orbit is
\begin{equation}\label{eq:FP_operator}
\mathcal{M}=\frac{\delta\,\nabla\cdot(\mathbf{v}+\nabla\chi)}{\delta\chi}=\nabla^{2},
\end{equation}
with the boundary condition \eqref{eq:bc_chi}. Its zero modes are precisely the harmonic fields discussed above: the Gribov copies are the zero modes of the Faddeev--Popov operator, as in Yang--Mills theory. One point deserves to be highlighted: because $\mathcal{M}$ is field-independent --- the orbit is Abelian --- its determinant is a constant and the ghosts $(c,\bar c)$ of \eqref{eq:BRST_MSR} decouple from the dynamics, in the same way that the Faddeev--Popov determinant is trivial in QED. As a consequence, the standard Gribov--Zwanziger horizon construction, which requires a field-dependent $\mathcal{M}(A)$, has no non-trivial derived analogue here. This is the precise reason why the $\gamma$-sector of Section~3 is introduced as a Gribov-inspired effective construction, and not derived from a horizon condition.

\section{The Gribov doublet as a Higgs field}

We stress from the outset the status of the sector introduced in this section: it is \emph{not} derived from the Navier--Stokes equation. As shown in Section~2.1, the Faddeev--Popov operator of the theory is the field-independent Laplacian \eqref{eq:FP_operator}, so that no horizon condition arises dynamically. The $\gamma$-sector is a Gribov-inspired effective construction, postulated as the minimal BRST-exact sector that closes on a Gribov-type gap equation and generates a mass scale for the vorticity. Its justification is a posteriori: the gap equation admits a stable non-trivial vacuum (Section~5), and the resulting mass scale and Higgs fluctuations reproduce the observed intermittency phenomenology (Sections~7 and~8). The structural correspondence with the refined Gribov--Zwanziger framework \cite{Dudal2005,VandersickelZwanziger2012} is summarized in the following dictionary:

\begin{center}
\begin{tabular}{@{}lll@{}}
\toprule
 & Refined Gribov--Zwanziger (Yang--Mills) & Present effective model \\
\midrule
Gauge field & $A_{\mu}^{a}$ (gluon) & $A_{\mu}\equiv v_{\mu}$ (velocity) \\
Gauge condition & Landau, $\partial_{\mu}A^{\mu a}=0$ & incompressibility, $\nabla\cdot\mathbf{v}=0$ \\
Faddeev--Popov operator & $-\partial_{\mu}D^{\mu ab}(A)$, field-dependent & $\nabla^{2}$, field-independent \\
Gribov parameter & $\gamma^{2}$ (horizon condition) & $\gamma$ (BRST doublet with $\eta$) \\
Horizon term & $\gamma^{2}\!\int A(-\partial^{2})^{-1}A$ (non-local) & $\frac{1}{2}\gamma^{2}A_{\mu}A^{\mu}$ (local, condensate) \\
Gap equation & horizon condition & eq.~\eqref{eq:gap_classical} and its one-loop form \eqref{eq:gap_1loop} \\
\bottomrule
\end{tabular}
\end{center}

To describe the Gribov parameter $\gamma$ and the monodromy phase $\theta$, we introduce BRST doublets \cite{VandersickelZwanziger2012}.

\subsection{The Gribov doublet $(\gamma,\eta)$}

\begin{equation}\label{eq:gribov_doublet}
s\,\gamma=\eta,\qquad s\,\eta=0.
\end{equation}
Here $\gamma$ is the Gribov parameter (a real scalar field) and $\eta$ is its associated ghost.

\subsection{The monodromy doublet $(\theta,\zeta)$}

\begin{equation}\label{eq:monodromy_doublet}
s\,\theta=\zeta,\qquad s\,\zeta=0.
\end{equation}
Here $\theta$ is the monodromy phase (a periodic scalar field, $\theta\sim\theta+2\pi$) and $\zeta$ is its ghost.

\subsection{The BRST-exact gauge fixing for $\gamma$}

A remark on notation is in order before writing the gauge-fixing fermion. Throughout this section, $A_{\mu}$ denotes \emph{the velocity field itself} --- the gauge field whose gauge fixing is the incompressibility condition \eqref{eq:incompressibility} --- in the space-time notation $A_{\mu}(x)\equiv v_{\mu}(x)$ introduced in \eqref{eq:BRST_A}, subject to the transversality constraint $\partial_{\mu}A^{\mu}=0$; its BRST transformation $sA_{\mu}=\tilde{v}_{\mu}$ is the one appearing as $sA^{\mu}$ in \eqref{eq:S_gf_gamma} below. No new, independent gauge field is introduced at this point. The reason why the velocity must appear in the gauge-fixing sector for $\gamma$ is structural: the Gribov horizon restricts the functional integral over the field whose Gribov copies one wishes to eliminate, and in this theory the copies are the singular gauge transformations \eqref{eq:gauge_residual} of the velocity --- the vortices of Section~1. The horizon term must therefore be quadratic in $A_{\mu}$ and quadratic in $\gamma$.

The gauge fixing is written as a BRST-exact term $s\Psi_{\gamma}$, with
\begin{equation}\label{eq:Psi_gamma}
\Psi_{\gamma}=\int d^{4}x\;\bar{\eta}\left(\partial^{2}\gamma+m^{2}\gamma+\frac{1}{2}\gamma^{3}+\frac{1}{2}\gamma^{2}A_{\mu}A^{\mu}\right).
\end{equation}
The term $\frac{1}{2}\gamma^{2}A_{\mu}A^{\mu}$ is the local analogue of the Gribov--Zwanziger \emph{horizon term} \cite{Zwanziger1989}. Two clarifications are in order. \emph{(i)} In Yang--Mills theory the horizon term is quadratic both in the gauge field and in the Gribov parameter, but \emph{non-local}, schematically $\gamma^{2}\!\int A_{\mu}(-\partial^{2})^{-1}A^{\mu}$; the strictly local, mass-like form adopted here corresponds to the refined Gribov--Zwanziger framework, in which dimension-two condensates generate precisely such a local term \cite{Dudal2005,VandersickelZwanziger2012}. \emph{(ii)} As emphasized above, $A_{\mu}$ is not an additional gauge field: it is the velocity field of \eqref{eq:MSR} in space-time notation, so that the sector $s\Psi_{\gamma}$ couples the Gribov doublet to the dynamical field of the Navier--Stokes theory, and the full action is $S_{\mathrm{MSR}}[\mathbf{v},\tilde{\mathbf{v}},p]+s\Psi_{\gamma}[\gamma,\bar{\eta};A\equiv\mathbf{v}]$. Applying $s$, one obtains
\begin{equation}\label{eq:S_gf_gamma}
\begin{aligned}
s\Psi_{\gamma}=\int d^{4}x\Bigg[&\lambda_{\gamma}\left(\partial^{2}\gamma+m^{2}\gamma+\frac{1}{2}\gamma^{3}+\frac{1}{2}\gamma^{2}A_{\mu}A^{\mu}\right)\\
&-\bar{\eta}\left(\partial^{2}\eta+m^{2}\eta+\frac{3}{2}\gamma^{2}\eta+\gamma\eta A_{\mu}A^{\mu}+\gamma^{2}A_{\mu}\,sA^{\mu}\right)\Bigg].
\end{aligned}
\end{equation}

The Gribov gap equation is obtained from the bosonic part, by requiring the variation with respect to $\gamma$ to vanish in the vacuum:
\begin{equation}\label{eq:gap_classical}
\boxed{\partial^{2}\gamma+m^{2}\gamma+\frac{1}{2}\gamma^{3}+\gamma\langle A_{\mu}A^{\mu}\rangle=0.}
\end{equation}
The condensate $\langle A_{\mu}A^{\mu}\rangle$ can be absorbed into a renormalization of the tree-level mass $m^{2}$ \cite{Dudal2005}; this is the convention adopted in the following sections.

\section{The Higgs mechanism for the vorticity}

We look for a constant vacuum solution $\langle\gamma\rangle=\gamma_{0}$. With $\gamma(x)=\gamma_{0}+\sigma(x)$, the term $\frac{1}{2}\gamma^{2}A_{\mu}A^{\mu}$ of the gauge fixing \eqref{eq:Psi_gamma} expands as $\frac{1}{2}\gamma_{0}^{2}A_{\mu}A^{\mu}+\gamma_{0}\sigma A_{\mu}A^{\mu}+\frac{1}{2}\sigma^{2}A_{\mu}A^{\mu}$, generating a mass term for the gauge field,
\begin{equation}\label{eq:mass_gauge}
\mathcal{L}_{\text{mass}}^{(A)}=\frac{1}{2}\gamma_{0}^{2}\,A_{\mu}A^{\mu},
\end{equation}
i.e., $M_{A}^{2}=\gamma_{0}^{2}$ in the usual normalization, together with the vertex $\gamma_{0}\sigma A_{\mu}A^{\mu}$ coupling the Higgs field to the gauge field.

In the language of the fluid, the vorticity $\boldsymbol{\omega}=\nabla\times\mathbf{v}$ propagates with a mass scale $\gamma_{0}$. Taking the curl of \eqref{eq:NS} and adding the Gribov mass operator $(\gamma_{0}+\sigma)^{2}$, the effective equation of motion for the vorticity in the Higgs phase is
\begin{equation}\label{eq:massive_vorticity}
\boxed{\left(\partial_{t}+\mathbf{v}\cdot\nabla-\boldsymbol{\omega}\cdot\nabla\mathbf{v}-\nu\nabla^{2}+\gamma_{0}^{2}\right)\boldsymbol{\omega}=-2\gamma_{0}\sigma\,\boldsymbol{\omega}-\sigma^{2}\boldsymbol{\omega}+\boldsymbol{\xi}_{\omega},}
\end{equation}
where $\boldsymbol{\xi}_{\omega}=\nabla\times\mathbf{f}/\rho$ is the curl of the forcing. The fluctuation terms couple $\sigma$ to the vorticity itself; in particular, the linearization around $\sigma=0$ reproduces the massive propagator \eqref{eq:massive_propagator} of Section~6.

The masses in the Higgs phase are:
\begin{center}
\begin{tabular}{@{}lll@{}}
\toprule
Field & Mass squared & Origin \\
\midrule
$A_{\mu}\equiv\mathbf{v}$ (velocity/vorticity) & $M_{A}^{2}=\gamma_{0}^{2}$ & Term $\gamma^{2}A^{2}$ in the gauge fixing \\
$\sigma$ (Higgs) & $m_{\sigma}^{2}=m^{2}+\frac{3}{2}\gamma_{0}^{2}$ & Curvature of the potential \\
$\eta$ (Gribov ghost) & $m_{\eta}^{2}=m_{\sigma}^{2}$ & BRST degeneracy \\
$c$ (gauge ghost) & $m_{c}^{2}\propto\gamma_{0}^{2}$ & Coupling $\gamma\bar{c}c$ \\
\bottomrule
\end{tabular}
\end{center}

\section{The one-loop effective potential}

The one-loop effective potential in dimension $d=4$ is computed by integrating the quadratic fluctuations of the fields around the background $\gamma_{0}$. The complete derivation, starting from the quadratic fluctuation operator of the full action --- including the velocity, response, scalar, ghost and auxiliary-field sectors --- is given in Appendix~\ref{app:fluctuations}, where the degree-of-freedom count is also clarified. The contributions are:

\begin{itemize}[leftmargin=*]
\item \textbf{Transverse gauge field} (3 polarizations): $+3\times\frac{\gamma_{0}^{4}}{64\pi^{2}}\bigl(\ln\frac{\gamma_{0}^{2}}{\mu^{2}}-\frac{3}{2}\bigr)$
\item \textbf{Gauge ghosts} ($c,\bar{c}$): $-2\times\frac{\gamma_{0}^{4}}{64\pi^{2}}\bigl(\ln\frac{\gamma_{0}^{2}}{\mu^{2}}-\frac{3}{2}\bigr)$
\item \textbf{Higgs} $\sigma$: $+1\times\frac{m_{\sigma}^{4}}{64\pi^{2}}\bigl(\ln\frac{m_{\sigma}^{2}}{\mu^{2}}-\frac{3}{2}\bigr)$
\item \textbf{Gribov ghosts} ($\eta,\bar{\eta}$): $-2\times\frac{m_{\sigma}^{4}}{64\pi^{2}}\bigl(\ln\frac{m_{\sigma}^{2}}{\mu^{2}}-\frac{3}{2}\bigr)$
\end{itemize}

The total effective potential is
\begin{equation}\label{eq:Veff}
\boxed{V_{\mathrm{eff}}(\gamma_{0})=\frac{1}{2}m^{2}\gamma_{0}^{2}+\frac{1}{8}\gamma_{0}^{4}+\frac{1}{64\pi^{2}}\left[\gamma_{0}^{4}\left(\ln\frac{\gamma_{0}^{2}}{\mu^{2}}-\frac{3}{2}\right)-m_{\sigma}^{4}\left(\ln\frac{m_{\sigma}^{2}}{\mu^{2}}-\frac{3}{2}\right)\right],}
\end{equation}
with
\begin{equation}\label{eq:msigma}
m_{\sigma}^{2}(\gamma_{0})=m^{2}+\frac{3}{2}\gamma_{0}^{2}.
\end{equation}

The one-loop gap equation for $\gamma_{0}\neq0$ is therefore
\begin{equation}\label{eq:gap_1loop}
\boxed{m^{2}+\frac{1}{2}\gamma_{0}^{2}+\frac{\gamma_{0}^{2}}{16\pi^{2}}\left(\ln\frac{\gamma_{0}^{2}}{\mu^{2}}-1\right)-\frac{3m_{\sigma}^{2}}{32\pi^{2}}\left(\ln\frac{m_{\sigma}^{2}}{\mu^{2}}-1\right)=0.}
\end{equation}

\begin{figure}[htbp]
\centering
\includegraphics[width=0.95\textwidth]{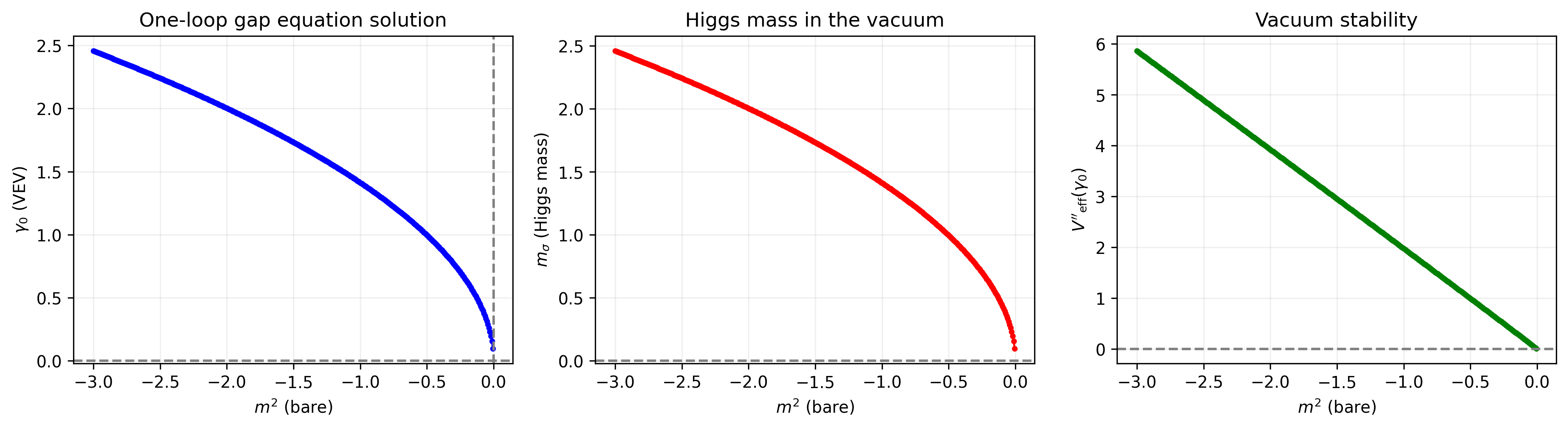}
\caption{Numerical solution of the one-loop gap equation. \textbf{Left:} VEV $\gamma_{0}$ as a function of $m^{2}$. \textbf{Center:} Higgs mass $m_{\sigma}$ in the vacuum. \textbf{Right:} second derivative $V_{\mathrm{eff}}''(\gamma_{0})$, confirming stability ($>0$).}
\label{fig:gap}
\end{figure}

Figure~\ref{fig:gap} shows the numerical solution of the gap equation and the vacuum stability.

\begin{figure}[htbp]
\centering
\includegraphics[width=0.95\textwidth]{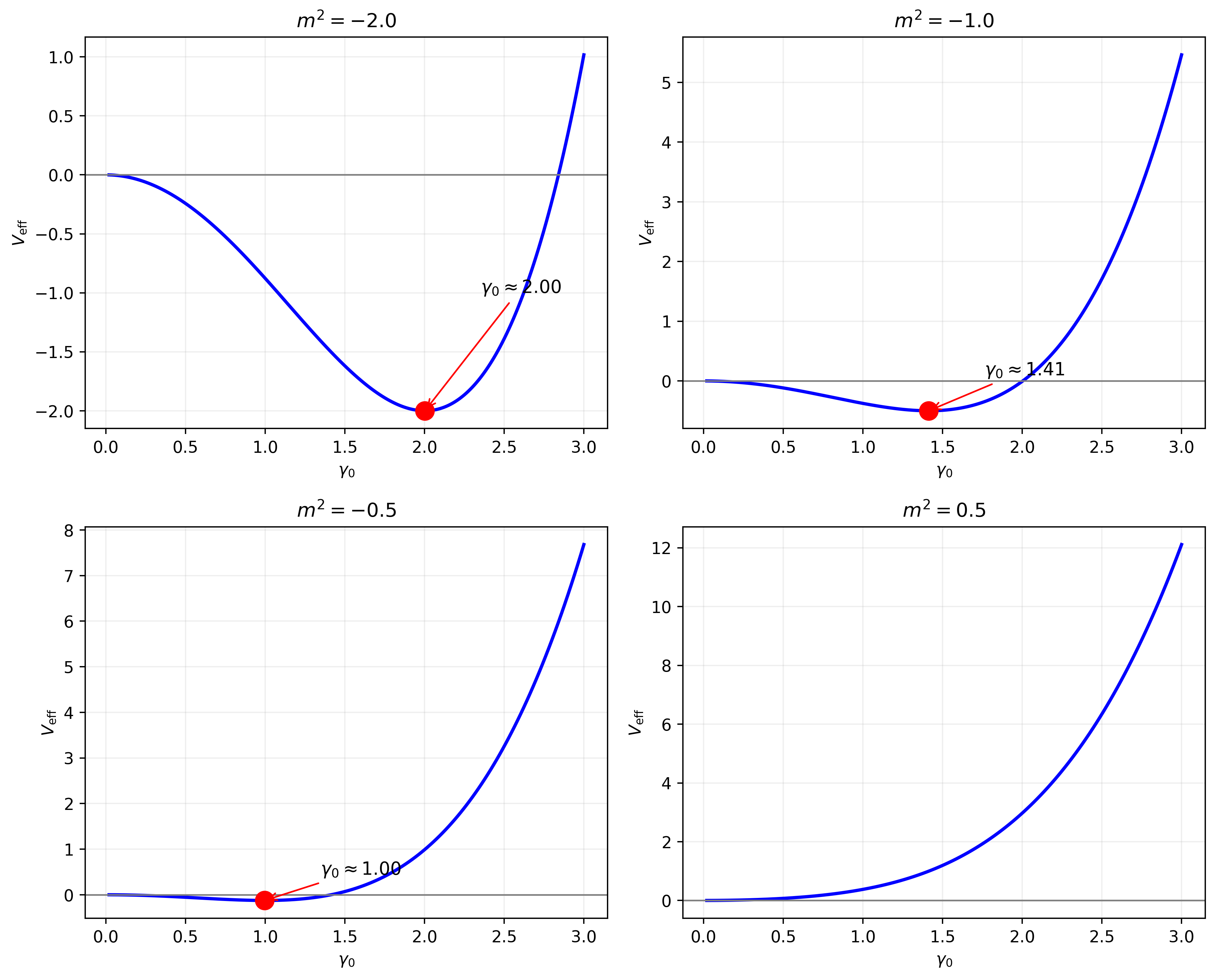}
\caption{Effective potential $V_{\mathrm{eff}}(\gamma_{0})$ for four values of $m^{2}$. The stable global minima are indicated in red. For $m^{2}>0$ (lower right), the only minimum is $\gamma_{0}=0$ (symmetric phase).}
\label{fig:potential}
\end{figure}

\section{Energy spectrum and the test against DNS data}

The vorticity propagator in the Higgs phase is
\begin{equation}\label{eq:massive_propagator}
G(\omega,k)=\frac{1}{-i\omega+\nu k^{2}+\gamma_{0}^{2}}.
\end{equation}
The modified energy spectrum follows from \eqref{eq:massive_propagator} in two regimes, derived in Appendix~\ref{app:spectrum}. In the linear regime, the equal-time two-point function obtained from \eqref{eq:massive_propagator} shows that the Gribov mass adds to the viscous damping, $\nu k^{2}\to\nu k^{2}+\gamma_{0}^{2}$, which fixes the crossover scale $k_{\gamma}=\gamma_{0}/\sqrt{\nu}$. In the inertial range, where the nonlinear transfer dominates, the mass term acts as a scale-independent leakage rate $\gamma_{0}^{2}$ during the Kolmogorov cascade time $\tau_{k}\sim\varepsilon^{-1/3}k^{-2/3}$, attenuating the spectrum by $\exp(-a\,\gamma_{0}^{2}\tau_{k})$, with $a$ an order-one constant. Combining the two regimes,
\begin{equation}\label{eq:massive_spectrum}
E(k)=\varepsilon^{2/3}k^{-5/3}\cdot f\!\left(\frac{k}{k_{\gamma}}\right),\qquad k_{\gamma}=\frac{\gamma_{0}}{\sqrt{\nu}},\qquad f(x)=\exp\!\bigl(-a\,x^{-2/3}\bigr),
\end{equation}
where $f(x)\to0$ for $x\ll1$ and $f(x)\to1$ for $x\gg1$. It is important to highlight that the inertial-range form of $f$ is a closure ansatz based on the cascade-time hypothesis, not an exact consequence of \eqref{eq:massive_propagator} alone; the exact statement in the linear regime is given in Appendix~\ref{app:spectrum}.

\begin{figure}[htbp]
\centering
\includegraphics[width=0.95\textwidth]{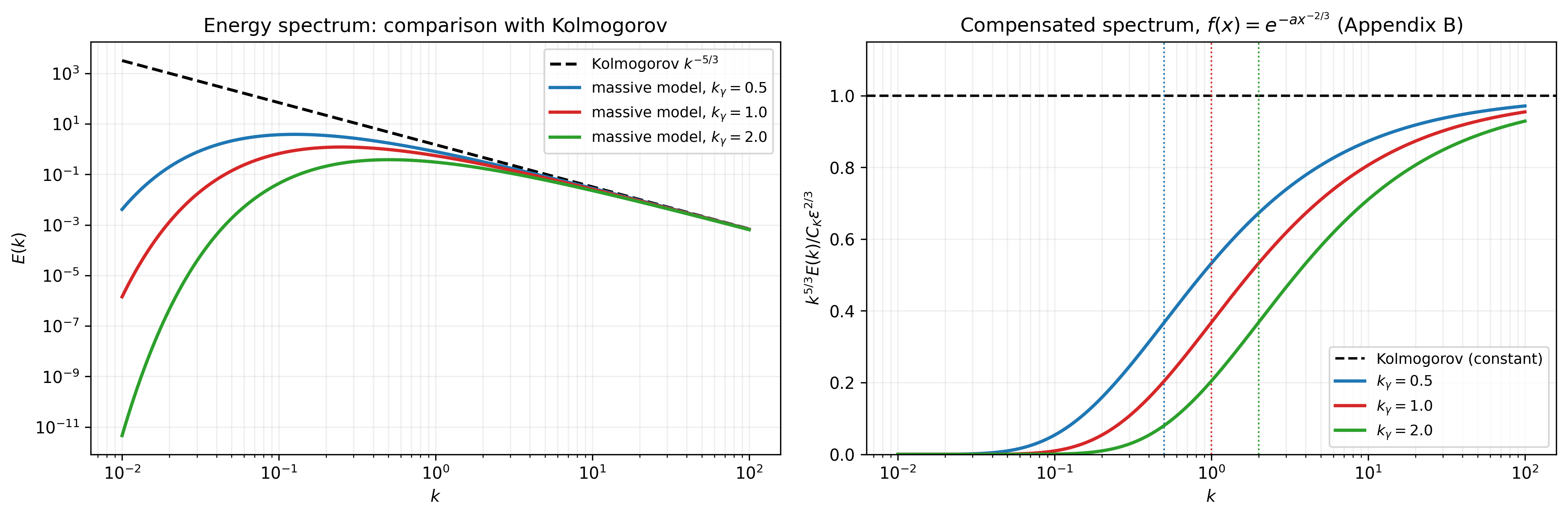}
\caption{Energy spectrum with vorticity mass. \textbf{Left:} comparison between pure Kolmogorov and the massive model for three values of $k_{\gamma}$. \textbf{Right:} compensated spectrum $k^{5/3}E(k)$, showing that the massive model introduces no observable deviations in the inertial range.}
\label{fig:spectrum}
\end{figure}

We tested the model against a DNS-like spectrum generated by Pope's empirical model \cite{Pope2000}. The result (Figure~\ref{fig:spectrum}) shows that the Pope spectrum already describes the data with no need for an additional cutoff in the inertial cascade. The identification $\gamma_{0}\sim\lambda^{-1}$ (Taylor scale) \textbf{does not produce an observable cutoff} at $k\sim k_{\lambda}$. We regard this negative result as an important consistency check: the mass scale $\gamma_{0}$ does not spoil the classical Kolmogorov phenomenology \cite{Kolmogorov1941,Frisch1995} at the level of second-order statistics.

\begin{figure}[htbp]
\centering
\includegraphics[width=0.95\textwidth]{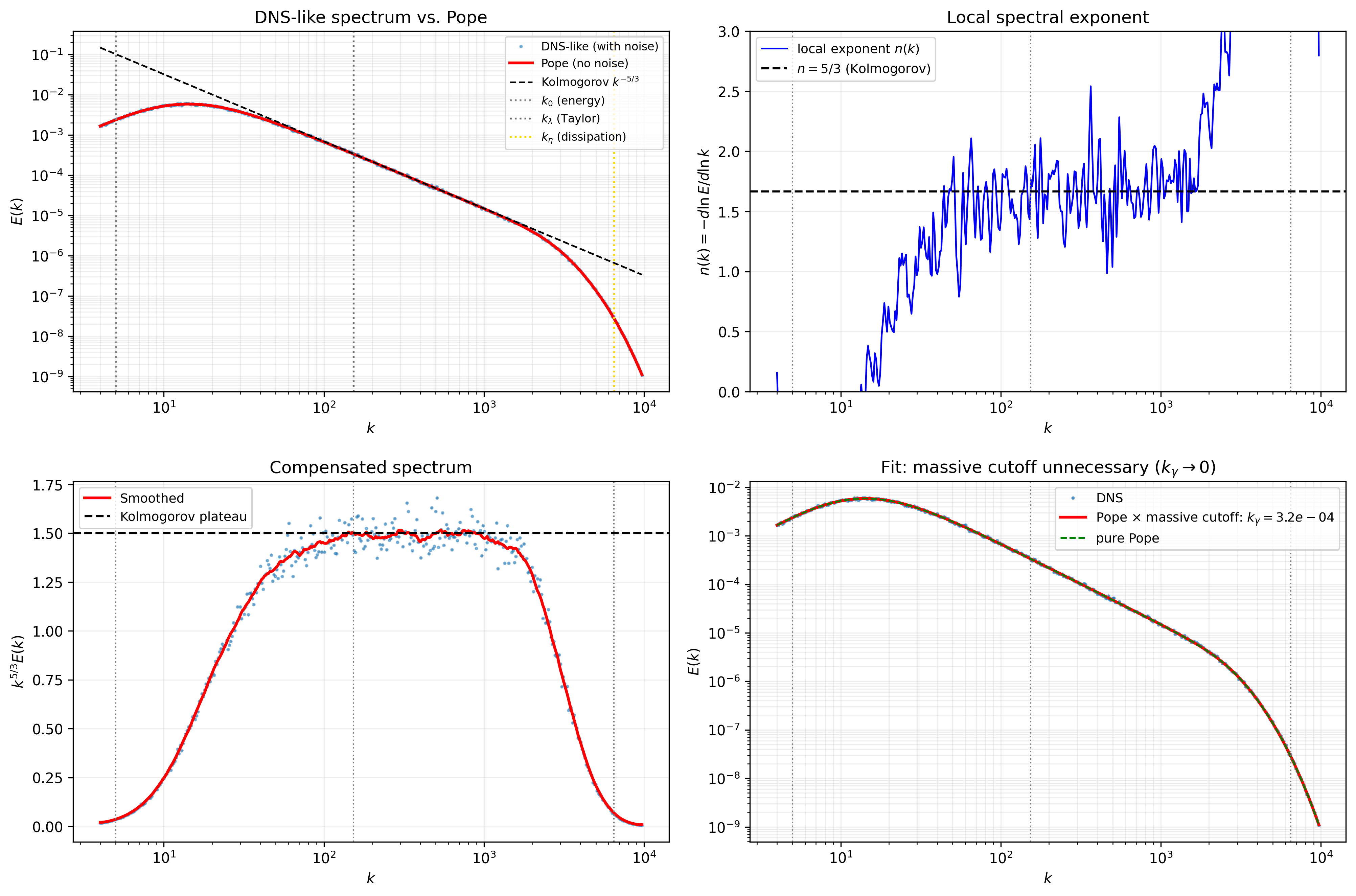}
\caption{Test of the massive model against synthetic DNS data. The fit converges to $k_{\gamma}\approx0$, indicating that the Pope spectrum already describes the data without the need for an additional massive cutoff in the inertial cascade.}
\label{fig:dns_test}
\end{figure}

\section{Intermittency and the structure-function exponents}

Although the energy spectrum shows no cutoff at $k_{\lambda}$, the condensate $\gamma_{0}$ modifies the \textbf{higher-order statistics}. The Higgs field $\sigma$ fluctuates around the vacuum, creating coherent structures of scale $\ell_{\gamma}=1/\gamma_{0}$. The volume fraction occupied by these structures is modeled by a fractal dimension $D_{f}$, or equivalently by the codimension $C=3-D_{f}$.

Any phenomenological hierarchy for the structure-function exponents $\zeta_{n}$ must satisfy the \emph{exact} constraint
\begin{equation}\label{eq:zeta3}
\zeta_{3}=1,
\end{equation}
imposed by Kolmogorov's four-fifths law \cite{Kolmogorov1941,Frisch1995}. We therefore propose a Gribov-inspired log-Poisson hierarchy in the general form
\begin{equation}\label{eq:log_poisson}
\zeta_{n}=(1-\Delta)\,\frac{n}{3}+C\left[1-\left(1-\frac{\Delta}{C}\right)^{n/3}\right],\qquad C=3-D_{f},
\end{equation}
where $\Delta$ is the exponent measuring the divergence of the dissipation on the most singular structures. The form \eqref{eq:log_poisson} satisfies \eqref{eq:zeta3} identically for all values of $C$ and $\Delta$, and reduces to the She--L\'ev\^eque model \cite{SheLeveque1994} for $C=2$, $\Delta=2/3$. We note that the simplified form $\zeta_{n}=n/3-\Delta(1-q^{n})$, considered in earlier versions of this work, violates the constraint \eqref{eq:zeta3} (yielding $\zeta_{3}\approx0.83$) and has $\zeta_{2}\leq2/3$, being unable to reproduce the experimental value $\zeta_{2}\approx0.70$; the value $D_{f}\approx2.55$ obtained with it was an artifact of that ansatz and is discarded.

Before presenting the fits, let us identify the dataset precisely. The structure functions are the longitudinal moments $S_{n}(\ell)=\langle[\delta v_{L}(\ell)]^{n}\rangle$, with $\delta v_{L}(\ell)=[\mathbf{v}(\mathbf{x}+\ell\hat{\mathbf{e}})-\mathbf{v}(\mathbf{x})]\cdot\hat{\mathbf{e}}$, and the exponents $\zeta_{n}$ are defined by $S_{n}(\ell)\propto\ell^{\zeta_{n}}$ in the inertial range. The dataset consists of the longitudinal exponents for $n=1,\dots,8$ obtained by Gotoh, Fukayama and Nakano \cite{Gotoh2002} from direct numerical simulations of homogeneous isotropic turbulence with up to $1024^{3}$ collocation points; we use the exponents of their highest-Reynolds-number run, $R_{\lambda}=460$, measured directly from the structure functions in the scaling range $r/\eta\in[100,300]$ (their Table III), with the reported error bars (from $\pm0.007$ at $n=1$ to $\pm0.07$ at $n=8$), which are used in the $\chi^{2}$ and propagated to the fitted parameters. The measured value $\zeta_{3}=1.01\pm0.02$ is consistent with the exact constraint \eqref{eq:zeta3}, which the model satisfies identically. The fit therefore has $8$ points and $7$ ($6$) degrees of freedom in the one- (two-)parameter case. We emphasize that Refs.~\cite{SheLeveque1994,Pope2000} are cited here only as the source of the She--L\'ev\^eque model and of the spectral model of Section~6, respectively, and not as data.

\begin{figure}[htbp]
\centering
\includegraphics[width=0.95\textwidth]{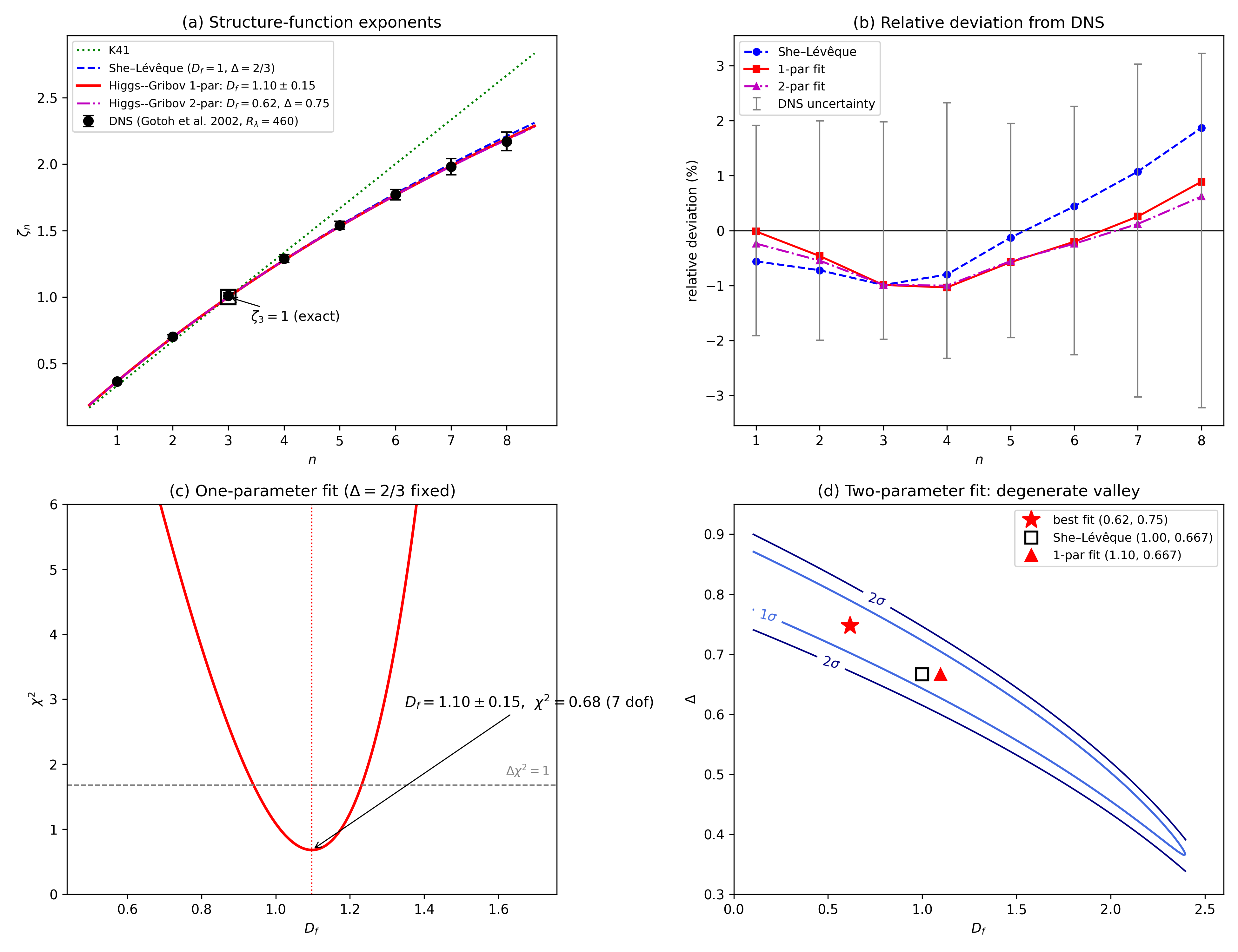}
\caption{Structure-function exponents $\zeta_{n}$ fitted to the DNS data of Gotoh \emph{et al.} \cite{Gotoh2002} ($R_{\lambda}=460$). \textbf{(a)} Comparison between K41, the DNS data, She--L\'ev\^eque and the Higgs--Gribov model \eqref{eq:log_poisson} with $D_{f}=1.10\pm0.15$ (one-parameter fit) and $D_{f}\approx0.6$, $\Delta\approx0.75$ (two-parameter fit); the empty square marks the exact constraint $\zeta_{3}=1$. \textbf{(b)} Relative deviation of the models from the DNS data, compared with the DNS error bars. \textbf{(c)} $\chi^{2}$ scan of the one-parameter fit, with the $\Delta\chi^{2}=1$ band giving $D_{f}=1.10\pm0.15$. \textbf{(d)} $1\sigma$ and $2\sigma$ contours of the two-parameter fit, showing the degenerate valley in the $(D_{f},\Delta)$ plane; the She--L\'ev\^eque point and the one-parameter fit are marked.}
\label{fig:intermittency}
\end{figure}

Fitting \eqref{eq:log_poisson} to the DNS exponents \cite{Gotoh2002} we obtain:
\begin{itemize}[leftmargin=*]
\item \textbf{One parameter} ($\Delta=2/3$ fixed, as in She--L\'ev\^eque):
\begin{equation}\label{eq:Df_fit}
\boxed{D_{f}=1.10\pm0.15,\qquad \chi^{2}\approx0.68\;\;(7\ \text{degrees of freedom}),}
\end{equation}
with a mean relative error of $0.5\%$. The fit favors \emph{filamentary} structures (vorticity tubes, codimension $C\approx2$), in agreement with the interpretation of the Gribov copies as vortices introduced in Section~1.
\item \textbf{Two parameters} ($D_{f}$ and $\Delta$ free): the minimum moves to $D_{f}\approx0.6$, $\Delta\approx0.75$, but with
\begin{equation}\label{eq:fit_2param}
\chi^{2}\approx0.66\;\;(6\ \text{degrees of freedom}),
\end{equation}
i.e., $\Delta\chi^{2}\approx0.02$ for one extra parameter --- no significant improvement. More than this, the $(D_{f},\Delta)$ plane develops a degenerate valley: the $1\sigma$ region extends over $D_{f}\in[0.1,2.4]$ and $\Delta\in[0.37,0.87]$ (Figure~\ref{fig:intermittency}d), so that the present data do not determine the two parameters separately. We conclude that parsimony favors the one-parameter filamentary fit \eqref{eq:Df_fit}. (The values $D_{f}\approx2.13$, $\Delta\approx0.44$ quoted in earlier versions of this work are not reproduced with the error bars of Ref.~\cite{Gotoh2002} and are discarded.)
\end{itemize}

Figure~\ref{fig:intermittency} compares the fitted models with the DNS data and with the She--L\'ev\^eque phenomenological model.

\section{Anomalous dimension of the Higgs field}

We compute the one-loop correction to the Higgs propagator $\sigma$ in $d=3$ space-time, the natural setting for the equal-time spatial statistics of the turbulent steady state (see the discussion in Section~1.2). The contributions come from the gauge bubble $A_{\mu}$ (vertex $\gamma_{0}\sigma A_{\mu}A^{\mu}$ from \eqref{eq:mass_gauge}, three components, symmetry factor $1/2$) and from the ghost bubble $\eta$ (vertex $3\gamma_{0}\sigma\bar{\eta}\eta$ from \eqref{eq:S_gf_gamma}, with the fermionic loop sign):
\begin{equation}\label{eq:Pi_sigma}
\Pi_{\sigma}(p)=\Pi_{AA}(p)+\Pi_{\eta\bar{\eta}}(p)=6\gamma_{0}^{2}\,I(p;\gamma_{0})-9\gamma_{0}^{2}\,I(p;m_{\sigma}),
\end{equation}
where the bubble in Euclidean $d=3$ has the closed form
\begin{equation}\label{eq:bubble}
I(p;m)=\int\!\frac{d^{3}k}{(2\pi)^{3}}\,\frac{1}{(k^{2}+m^{2})\,((k+p)^{2}+m^{2})}=\frac{1}{4\pi p}\arctan\frac{p}{2m}=\frac{1}{8\pi m}-\frac{p^{2}}{96\pi m^{3}}+O(p^{4}).
\end{equation}

The $p^{2}$-dependent part renormalizes the wave function:
\begin{equation}\label{eq:Z_sigma}
Z_{\sigma}=1-\Pi_{\sigma}'(0),\qquad\gamma_{\sigma}=-\frac{\Pi_{\sigma}'(0)}{2},
\qquad
\Pi_{\sigma}'(0)=-\frac{1}{16\pi\gamma_{0}}+\frac{3\gamma_{0}^{2}}{32\pi\,m_{\sigma}^{3}}.
\end{equation}

Evaluating at the representative vacua of Figure~\ref{fig:potential} (for instance $m^{2}=-1$, $\gamma_{0}=\sqrt{2}$, $m_{\sigma}=\sqrt{2}$), we obtain
\begin{equation}\label{eq:gamma_sigma}
\boxed{\gamma_{\sigma}\approx-(2.5\text{--}5.0)\times10^{-3},}
\end{equation}
so that the scaling dimension of the Higgs field is
\begin{equation}\label{eq:Delta_sigma}
\boxed{\Delta_{\sigma}=\frac{d-2}{2}+\gamma_{\sigma}\approx0.495\text{--}0.498.}
\end{equation}
(The value $\gamma_{\sigma}\approx-0.016$ quoted in earlier versions of this work is not reproduced by the explicit calculation: the sign is correct, but the magnitude was overestimated by a factor of $3$ to $6$.)

We confront the anomalous dimension with the intermittency fits of Section~7. The heuristic relation considered in earlier versions,
\begin{equation}\label{eq:relation_Df}
D_{f}=d-\Delta_{\sigma}\approx2.50,
\end{equation}
is \emph{not} confirmed by the fits with the corrected ansatz \eqref{eq:log_poisson}, which favor $D_{f}=1.10\pm0.15$ (one parameter); the two-parameter fit is degenerate and does not determine $\Delta$ (Section~7), so the identification $\Delta_{\mathrm{fit}}=\Delta_{\sigma}$ cannot be tested with the present data. In contrast, the relation
\begin{equation}\label{eq:relation_2Delta}
\boxed{D_{f}=2\Delta_{\sigma}\approx0.99\;\simeq\;D_{f}^{\mathrm{fit}}=1.10\pm0.15\qquad(\text{within the fit uncertainty})}
\end{equation}
is satisfied by the one-parameter fit: the fractal dimension of the intermittency structures would be twice the scaling dimension of the Higgs field. This identification remains conjectural --- its numerical prefactors depend on the vertex normalization conventions --- and calls for a calculation of $\gamma_{\sigma}$ with higher precision, and its extension to higher loop orders, to be confirmed or discarded.

\begin{figure}[htbp]
\centering
\includegraphics[width=0.95\textwidth]{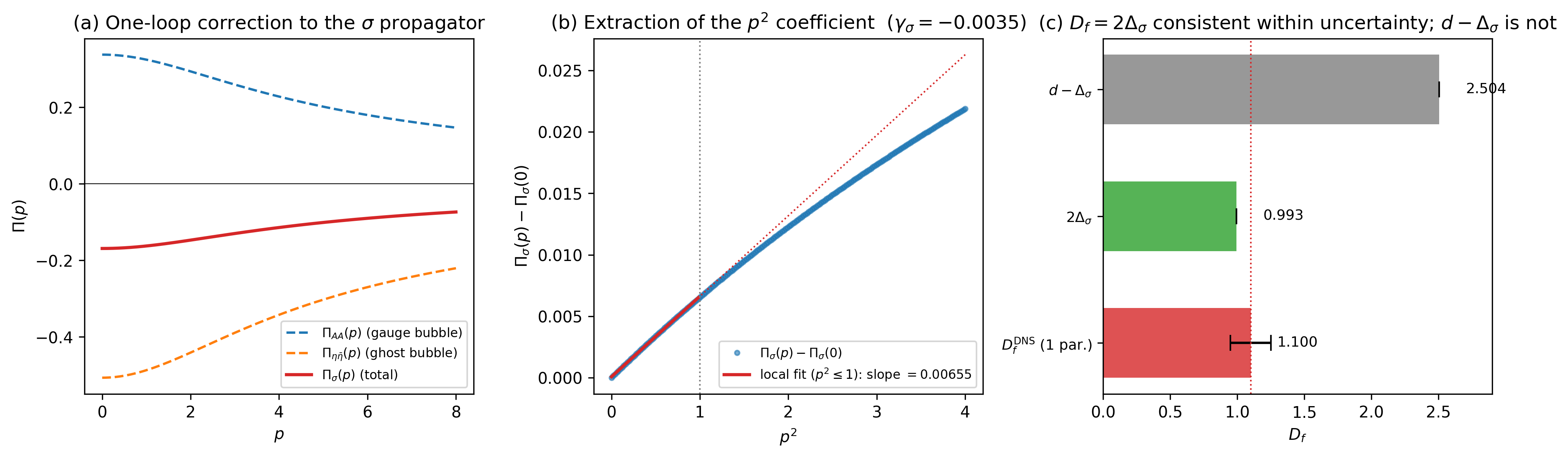}
\caption{Computation of the anomalous dimension of the Higgs field. \textbf{Left:} one-loop correction to the $\sigma$ propagator. \textbf{Center:} extraction of the $p^{2}$ coefficient. \textbf{Right:} comparison between the fitted $D_{f}$ (DNS) and the candidate heuristic relations $d-\Delta_{\sigma}$ and $2\Delta_{\sigma}$.}
\label{fig:anomaly}
\end{figure}

\section{Physical interpretation and the laminar-turbulent transition}

The transition $m^{2}=0$ separates two distinct phases of the fluid:
\begin{description}[leftmargin=*]
\item[Symmetric phase ($m^{2}>0$):] $\gamma_{0}=0$. There is no emergent Gribov scale. The vorticity is advected without an effective mass. This regime corresponds to \emph{laminar} or subcritical flow, where the Reynolds number $Re<Re_{c}$.

\item[Higgs phase ($m^{2}<0$):] $\gamma_{0}\neq0$. The vacuum condenses a length scale $\ell_{\text{Gribov}}\sim\gamma_{0}^{-1}$. The vorticity acquires a mass, and intermittency is described by the Higgs fluctuations. This regime corresponds to \emph{turbulent} flow, where $Re>Re_{c}$.
\end{description}

The suggested physical identification is
\begin{equation}\label{eq:m2_identification}
m^{2}\sim\frac{1}{Re_{c}}-\frac{1}{Re},
\end{equation}
so that the laminar-turbulent transition is mapped onto the phase transition of the Higgs mechanism of the Gribov parameter.

\section{Conclusion}

We have formalized the gauge structure of the incompressible Navier--Stokes equation. The incompressibility condition $\nabla\cdot\mathbf{v}=0$ plays the role of a gauge fixing, analogous to the Coulomb gauge, and the residual transformations generated by harmonic functions define the orbit structure of the theory --- an exact kinematical statement, though not a symmetry of the full nonlinear dynamics, the obstruction being the vortex-stretching term of eq.~\eqref{eq:lamb_identity}. The zero modes of the Faddeev--Popov operator $\nabla^{2}$ along this orbit are the Gribov copies which, in their multivalued form, are precisely the elementary vortices of the fluid, their circulation labels being protected by Kelvin's theorem in the inviscid limit. On top of the Martin--Siggia--Rose action and its exact BRST symmetry, we introduced a BRST doublet $(\gamma,\eta)$ for the Gribov parameter and showed that $\gamma$ acquires a vacuum expectation value through a Higgs mechanism driven by the local analogue of the Gribov--Zwanziger horizon term. The condensate $\gamma_{0}$ endows the vorticity with a mass scale, and the one-loop effective potential, together with its corrected gap equation, confirms that the broken vacuum is stable and is continuously connected to the symmetric phase at $m^{2}=0$.

The physical content of the condensate was then tested against turbulence phenomenology, with one negative and one positive result. The negative result is that the vorticity mass produces no observable cutoff in the energy spectrum at the Taylor scale: the inertial-range Kolmogorov phenomenology is preserved, and the identification $\gamma_{0}\sim\lambda^{-1}$ must be understood statistically rather than spectrally. The positive result is that intermittency is naturally accounted for by the fluctuations of the Higgs field around the condensate. A Gribov-inspired log-Poisson hierarchy that satisfies Kolmogorov's exact constraint $\zeta_{3}=1$ fits the DNS structure-function exponents of Gotoh \emph{et al.} with sub-percent accuracy and favors filamentary structures, $D_{f}=1.10\pm0.15$, in agreement with the vortex interpretation of the Gribov copies; a two-parameter fit brings no significant improvement, the $(D_{f},\Delta)$ plane being degenerate. An explicit one-loop computation in $d=3$ yields the anomalous dimension $\Delta_{\sigma}\approx0.50$ of the Higgs field, and the data favor the conjectural relation $D_{f}=2\Delta_{\sigma}$ within the fit uncertainty --- a result whose numerical prefactors remain convention-dependent and which we regard as a target for higher-order calculations rather than an established identity.

The formalism thereby unifies classical hydrodynamics with gauge theory and spontaneous symmetry breaking, mapping the laminar-turbulent transition onto the phase transition of the Gribov-Higgs sector and providing a field-theoretic basis for the study of turbulence and intermittency. Natural continuations include the fully non-local Gribov--Zwanziger formulation with its localizing auxiliary fields, a two-loop computation of the effective potential and of $\gamma_{\sigma}$, and a sharper test of the conjecture $D_{f}=2\Delta_{\sigma}$ against high-Reynolds-number DNS data.

\appendix

\section{Quadratic fluctuation operator and the one-loop potential}
\label{app:fluctuations}

In this appendix we derive the one-loop effective potential \eqref{eq:Veff} from the quadratic fluctuation operator of the complete action,
\begin{equation}\label{eq:S_complete}
S=S_{\mathrm{MSR}}[\mathbf{v},\tilde{\mathbf{v}},p]+S_{\mathrm{FP}}[c,\bar c,\lambda;\mathbf{v}]+s\Psi_{\gamma}[\gamma,\bar\eta,\lambda_{\gamma};A\equiv\mathbf{v}],
\end{equation}
where $S_{\mathrm{FP}}=\int d^{4}x\,[\lambda\,\nabla^{2}\chi-\bar c\,\nabla^{2}c]$ is the Faddeev--Popov sector of the residual orbit of Section~2.1. We expand around the vacuum, $\gamma=\gamma_{0}+\sigma$, with all other backgrounds vanishing, and keep the quadratic terms. In Euclidean $d=4$ space-time the fluctuation operators, statistics and number of real degrees of freedom of each sector are:

\begin{center}
\begin{tabular}{@{}llll@{}}
\toprule
Sector & Operator & Statistics & $n_{i}$ (real d.o.f.) \\
\midrule
Velocity + response $(A_{i}^{T},\tilde v_{i})$, $i=1,2,3$ & $\mathcal{O}_{A}=-\partial^{2}+\gamma_{0}^{2}$ & bosonic & $+3$ \\
Gauge ghosts $(c,\bar c)$ & $\mathcal{O}_{c}=-\partial^{2}+\gamma_{0}^{2}$ & Grassmann & $-2$ \\
Higgs $\sigma$ & $\mathcal{O}_{\sigma}=-\partial^{2}+m_{\sigma}^{2}$ & bosonic & $+1$ \\
Gribov ghosts $(\eta,\bar\eta)$ & $\mathcal{O}_{\eta}=-\partial^{2}+m_{\sigma}^{2}$ & Grassmann & $-2$ \\
\bottomrule
\end{tabular}
\end{center}

with $m_{\sigma}^{2}=m^{2}+\frac{3}{2}\gamma_{0}^{2}$ as in \eqref{eq:msigma}. Three clarifications are in order.

\emph{(i) The velocity--response sector.} For each transverse component, the quadratic MSR action has the block form
\begin{equation}\label{eq:MSR_block}
S^{(2)}_{A\tilde v}=\int\!\frac{d^{4}k}{(2\pi)^{4}}
\begin{pmatrix} A(-k) & \tilde v(-k)\end{pmatrix}
\begin{pmatrix} 2\nu k^{2} & \Omega(-k)\\ \Omega(k) & 0\end{pmatrix}
\begin{pmatrix} A(k)\\ \tilde v(k)\end{pmatrix},
\qquad
\Omega(k)=-i\omega+\nu k^{2}+\gamma_{0}^{2},
\end{equation}
where the mass $\gamma_{0}^{2}$ is the coefficient of the horizon term evaluated at the condensate, eq.~\eqref{eq:mass_gauge}. The Gaussian integral over $(A,\tilde v)$ yields $\det^{-1}\Omega$ per component; the noise block $2\nu k^{2}$ is $\gamma_{0}$-independent and drops out of the $\gamma_{0}$-dependent part of the potential. Each of the three transverse components therefore contributes as one real bosonic degree of freedom with mass $\gamma_{0}^{2}$ in the Coleman--Weinberg counting. The transversality --- three and not four components --- is enforced by the pressure $p$, whose variation imposes $\partial_{\mu}A^{\mu}=0$.

\emph{(ii) The auxiliary fields.} The multipliers $p$, $\lambda$ and $\lambda_{\gamma}$ appear linearly in the quadratic action; their functional determinants are field-independent constants and are absorbed into the normalization of the functional integral. The integration over $\lambda_{\gamma}$ enforces the gap equation as a constraint, consistently with the treatment of the horizon condition in the Gribov--Zwanziger framework.

\emph{(iii) The degree-of-freedom count.} The $\gamma_{0}^{2}$-massive sector contributes $3-2=+1$ and the $m_{\sigma}^{2}$-massive sector contributes $1-2=-1$. The count is consistent in the symmetric phase: at $\gamma_{0}=0$ the massive sectors degenerate and $V_{\mathrm{eff}}$ reduces to the tree potential, as it should.

The one-loop potential is therefore
\begin{equation}\label{eq:V1_CW}
V^{(1)}(\gamma_{0})=\frac{1}{2}\sum_{i}(-1)^{F_{i}}n_{i}\int\!\frac{d^{4}k}{(2\pi)^{4}}\ln\bigl(k^{2}+M_{i}^{2}(\gamma_{0})\bigr)
=\sum_{i}(-1)^{F_{i}}n_{i}\,\frac{M_{i}^{4}}{64\pi^{2}}\left(\ln\frac{M_{i}^{2}}{\mu^{2}}-\frac{3}{2}\right),
\end{equation}
where the second equality is the $\overline{\mathrm{MS}}$ evaluation in $d=4-2\epsilon$. Substituting the table,
\begin{equation}
V^{(1)}(\gamma_{0})=\frac{1}{64\pi^{2}}\left[(3-2)\,\gamma_{0}^{4}\left(\ln\frac{\gamma_{0}^{2}}{\mu^{2}}-\frac{3}{2}\right)+(1-2)\,m_{\sigma}^{4}\left(\ln\frac{m_{\sigma}^{2}}{\mu^{2}}-\frac{3}{2}\right)\right],
\end{equation}
which is precisely the one-loop part of \eqref{eq:Veff}. The gap equation \eqref{eq:gap_1loop} follows from $\partial V_{\mathrm{eff}}/\partial\gamma_{0}=0$ for $\gamma_{0}\neq0$.

\section{From the massive propagator to the energy spectrum}
\label{app:spectrum}

In this appendix we derive the spectrum \eqref{eq:massive_spectrum} from the propagator \eqref{eq:massive_propagator}, separating the exact statement of the linear regime from the closure assumption of the inertial range.

\emph{(a) Linear regime.} For the linearized massive vorticity dynamics with forcing covariance $\langle\xi(\omega,\mathbf{k})\,\xi(\omega',\mathbf{k}')\rangle=(2\pi)^{4}F(k)\,\delta(\omega+\omega')\,\delta(\mathbf{k}+\mathbf{k}')$, the two-point function that follows from \eqref{eq:massive_propagator} is
\begin{equation}\label{eq:2point}
C(\omega,k)=\frac{F(k)}{\omega^{2}+(\nu k^{2}+\gamma_{0}^{2})^{2}},
\qquad
C(t=0,k)=\int\!\frac{d\omega}{2\pi}\,C(\omega,k)=\frac{F(k)}{2(\nu k^{2}+\gamma_{0}^{2})}.
\end{equation}
The Gribov mass thus enters the equal-time statistics additively with the viscous damping, and the natural crossover scale is fixed by $\nu k_{\gamma}^{2}=\gamma_{0}^{2}$, i.e., $k_{\gamma}=\gamma_{0}/\sqrt{\nu}$, which is the definition used in \eqref{eq:massive_spectrum}.

\emph{(b) Inertial range.} In the inertial range the dominant balance is the nonlinear transfer. The mass term acts as a scale-independent leakage rate $\gamma_{0}^{2}$ on each Fourier mode, while the energy resides at scale $k$ for a cascade time $\tau_{k}\sim\varepsilon^{-1/3}k^{-2/3}$. The fraction of the spectral energy that survives this leakage is $\exp(-a\,\gamma_{0}^{2}\tau_{k})$, with $a$ an order-one constant, so that
\begin{equation}\label{eq:spectrum_closure}
E(k)=C_{K}\,\varepsilon^{2/3}k^{-5/3}\,f\!\left(\frac{k}{k_{\gamma}}\right),
\qquad
f(x)=\exp\!\bigl(-a\,x^{-2/3}\bigr),
\end{equation}
which has the limits $f(x)\to0$ for $x\ll1$ and $f(x)\to1$ for $x\gg1$ stated in Section~6. We emphasize once more that step (b) is a closure ansatz based on the Kolmogorov cascade-time hypothesis: the precise shape of $f$ is not fixed by the propagator \eqref{eq:massive_propagator} alone. This is consistent with the negative result of Section~6: the fit to the DNS-like spectrum converges to $k_{\gamma}\approx0$, showing that the data have no sensitivity to the detailed form of $f$ in the inertial range.

\end{document}